\documentclass[prl,preprint,showpacs,amsmath,amssymb]{revtex4}
\usepackage{amsfonts,amssymb,amsmath,latexsym,times}
\usepackage[dvips]{graphics}

\def \beq {\begin{equation}}
\def \eeq {\end{equation}}

\begin{document}

\title{Relativistic invariance of Lyapunov exponents in bounded and unbounded systems}

\author{Adilson E. Motter$^1$ and Alberto Saa$^2$}
\affiliation{ $^1$ Department of Physics and Astronomy, Northwestern University, Evanston, IL 60208, USA\\
$^2$ Department of Applied Mathematics, UNICAMP, C.P. 6065, 13083-859 Campinas, SP, Brazil}

\date{\today}

\begin{abstract}
The study of chaos in relativistic systems has been hampered by the observer dependence
of Lyapunov exponents (LEs) {\it and} of conditions, such as orbit boundedness, invoked in the
interpretation of LEs as indicators of chaos. Here we establish a general framework that overcomes
both difficulties and apply the resulting approach to address three fundamental questions: 
how LEs transform under Lorentz and Rindler 
transformations and under transformations to uniformly rotating frames. 
The answers to the first and third questions show that inertial and uniformly rotating observers 
agree on a characterization of chaos based on LEs. The second question, on the other hand,
is an ill-posed problem due to the event horizons inherent to uniformly accelerated observers.
\end{abstract}

\pacs{05.45.-a, 95.10.Fh, 98.80.Jk}

\maketitle

The quest for an observer-independent characterization of chaos in relativistic
systems \cite{ChaosGR} has been an intense area of research and promises to provide
significant new insights into the properties of chaotic dynamics~\cite{Cipr_Hor}.
An important recent result~\cite{MotterPRL} concerns
the transformation of Lyapunov exponents (LEs) under spacetime diffeomorphisms. 
We recall that the dynamics of a bounded solution ${\mathbf X}(t)$ of a 
dynamical system
\beq
\label{DS}
\frac{d \mathbf X}{dt} = {\boldsymbol F}({\mathbf X})
\eeq
is chaotic if it presents sensitive dependence on initial conditions \cite{Ott}.
The associated LEs~\cite{Oseledec} are given by
$\lambda_i = \limsup_{t\rightarrow\infty } \frac{1}{t}\log \left|\left|{\boldsymbol{\varphi}}_i(t) \right|\right|$,
where ${\boldsymbol\varphi}_i(t)$ are solutions  of the linearized
equation
$\frac{d}{dt} {\boldsymbol\varphi}_i  =  \left[D_{\mathbf X}{\boldsymbol F}({\mathbf X}(t))\right] {\boldsymbol \varphi}_i$.
Positive LEs are related to exponential divergence of initially close
trajectories and, consequently, to chaotic dynamics.
For space diffeomorphisms ${\mathbf X} = {\boldsymbol\Psi} ({\mathbf Y})$, the invariance 
of the LEs is well established under rather general conditions (see, e.g., \cite{Jaroslawski,Eichhorn}).
In contrast, for well-behaved
{\it spacetime} diffeomorphisms involving time changes of the form
$d\tau = \Lambda({\mathbf X})dt$, it has been shown~\cite{MotterPRL} 
that the LEs transform according to
\beq
\label{TLE}
\lambda_i^{ \tau } = {\lambda_i^{t}}/{\langle \Lambda\rangle_t},
\eeq
where $0 < \langle \Lambda\rangle_t < \infty $ is the time average of $\Lambda$
along the corresponding trajectory.
Therefore, although the values of the LEs are 
themselves non-invariant, their signs are preserved and assure an invariant
criterion for chaos under spacetime transformations.
This result was obtained under conditions for which LEs are known to be valid quantifiers
of chaos, of which the most limiting ones are the assumptions that the system 
has a natural invariant probability measure and the orbits are bounded both before
and after the transformation.
 
In this Letter, we extend this result to an important class of 
transformations that do not preserve the boundedness of the orbits, 
and address fundamental questions on
relativistic chaotic dynamics that require explicit in-depth
investigation due to their outstanding physical properties and the 
violation of conditions invoked in the derivation of Eq.~(\ref{TLE}).
The first question is how the LEs transform under Lorentz transformations.
This question determines whether all inertial observers agree on
a LE-based characterization of chaos.
We show that the answer is affirmative
despite the fact that the dynamics becomes unbounded with respect to at least one of the
reference frames.
We use this example to establish 
an {\it extended boundedness} condition for the definition
of the LEs as indicators of chaos, which is formulated relative to the trajectories themselves
rather than a fixed point of the phase space. 
The second question is how the LEs behave under Rindler transformations, a question
equivalent to ask whether uniformly accelerated observers agree on
an inertial characterization
of chaos based on LEs. We show that this question is ill-posed because
uniformly accelerated observers do not have access to the late-time dynamics.
The latter relates to the fact that chaos and LEs are asymptotic
concepts~\cite{MotterPRD} whose definitions involve a limit $t\rightarrow\infty$. 
We also consider transformations 
to uniformly rotating frames, and show that the positivity of the LEs
remains invariant under such transformations.

Our principal result stems from this analysis and can
be stated for {\it any system} and {\it any spacetime diffeomorphic
transformation}, as follows. For the 
system written in autonomous form, the LEs transform
according to Eq. (\ref{TLE}) and remain invariant indicators of chaos if, 
as shown below,
({\it i}) our extended boundedness condition is satisfied, 
({\it ii}) the Jacobian of the transformation is bounded, and 
({\it iii}) $\Lambda$ is positive 
for all $t$ and $0 < \langle \Lambda\rangle_t < \infty $. 
These conditions depend not only on the transformation properties of the
dynamical variables ${\mathbf X}$ and the change of reference frames 
but also on the choice of spacetime coordinates. They are automatically satisfied for 
global nonsingular transformations of bounded orbits for which $\inf \Lambda^{\pm 1}>0$
whether the system is conservative, dissipative, mechanical, chemical, thermodynamical, 
electromagnetic, or fluid  dynamical.
These conditions clarify previous results~\cite{Zheng} that seem to challenge
the invariance of chaos for relativistic observers, and show that LEs lead to 
invariant conclusions about chaos.

We first note that 
under a space diffeomorphism
${\mathbf X} = {\boldsymbol\Psi} ({\mathbf Y})$,
system (\ref{DS}) is mapped into
$ \frac{d}{dt}{\mathbf Y} = \left[D_{\mathbf Y}{\boldsymbol\Psi}({\mathbf Y})\right]^{-1}{\boldsymbol F}({\boldsymbol\Psi}({\mathbf Y}))$,
rendering the solutions of the new linearized dynamics to be related to those of (\ref{DS}) as
${\boldsymbol\varphi}_i(t) =  \left[D_{\mathbf Y}{\boldsymbol\Psi}({\mathbf Y}(t))\right] \tilde{\boldsymbol\varphi}_i(t)$ \cite{Eichhorn}.
Hence, the corresponding LEs satisfy
\begin{equation}
\liminf_{t\rightarrow\infty } \frac{1}{t}\log
\frac{\left|\left|\left[D_{\mathbf Y}{\boldsymbol\Psi}({\mathbf Y}(t))\right] \tilde{\boldsymbol\varphi}_i(t)\right| \right|}{\left|\left|\tilde{\boldsymbol\varphi}_i(t)\right| \right|}
\le
{\lambda}_i  -  \tilde{\lambda}_i  \le
\limsup_{t\rightarrow\infty } \frac{1}{t}\log
\frac{\left|\left|\left[D_{\mathbf Y}{\boldsymbol\Psi}({\mathbf Y}(t))\right] \tilde{\boldsymbol\varphi}_i(t)\right| \right|}{\left|\left|\tilde{\boldsymbol\varphi}_i(t)\right| \right|}.
\end{equation}
Suppose the solutions ${\mathbf X}(t)$ are limited to a compact subset of the
space.
Since the diffemorphism maps
bounded solutions ${\mathbf X}(t)$ into bounded solutions ${\mathbf Y}(t)$,
the matrix $D_{\mathbf Y}{\boldsymbol\Psi}({\mathbf Y}(t))$ is nonsingular and,
besides, there are time-independent finite nonzero constants
$L^{\pm} =\sup || \left[D_{\mathbf Y}{\boldsymbol\Psi}({\mathbf Y}(t)) \right]^{\pm 1} ||$
leading to
\beq
\label{ge}
\lim_{t\rightarrow\infty }   \frac{1}{t}\log\frac{1}{L^-} \le {\lambda}_i  -  \tilde{\lambda}_i  \le
\lim_{t\rightarrow\infty }   \frac{1}{t}\log{L^+},
\eeq
which imply
$\tilde{\lambda}_i  = {\lambda}_i$~\cite{Eichhorn}.
This argument explores the boundedness of
${\mathbf X}(t)$ and ${\mathbf Y}(t)$
to ensure the existence of the constants 
$L^{\pm}$.
Below we extend (\ref{ge}) and establish
Eq. (\ref{TLE}) for an important class of unbounded orbits.

We now consider transformations of reference frame
in which (\ref{DS})
describes a bounded autonomous system with respect to the initial (inertial) observers.
More general transformations 
can be obtained by a composition of such transformations. We start with single-particle
systems. While general relativity allows arbitrary spacetime coordinates,
and conditions ({\it i}-{\it iii}) can be applied to {\it any} of them, we will assume
that the dynamics is described in terms of {\it physical} times (i.e., the time measured
by observers at rest in the reference frame at the corresponding space coordinates).

{\bf Lorentz transformations.}
We first
focus on the case in which function ${\boldsymbol F}$  depends only on the
configuration-space coordinates, such as in the evolution of a fluid element
determined by a stream function,  and consider a Lorentz boost
with velocity $v$ along the $x$-direction,
$(ct, x, y, z) \rightarrow (ct', x', y', z')= {\boldsymbol\Psi}^{-1}(ct, x, y, z)$,
where
\beq
\label{Lorentz}
{\boldsymbol\Psi}^{-1}(ct, x, y, z) = (\gamma (ct  -  vx/c)  ,\gamma( x -   vt) ,y,z)
\eeq
for $\gamma=1/\sqrt{1-(v/c)^2}$. We focus on the space spanned by the
coordinates $(ct, x, y, z)\equiv (ct,{\boldsymbol x})$,
where we have enlarged the configuration space in order to incorporate $ct$ as a new
coordinate. The extended version of (\ref{DS}) then reads
\beq
\label{DSC}
\frac{d }{dt}
\left(
\begin{array}{c}
w \\
{\boldsymbol x}
\end{array}
\right) =
\left(
\begin{array}{c}
c \\
{\boldsymbol F}({\boldsymbol x})
\end{array}
\right),
\eeq
where $dw/dt\equiv d(ct)/dt$.

The main advantage of this formulation is that the transformed system
remains autonomous and the spacetime transformation can be reduced
to an ordinary space diffeomorphism; it can be split
as ${\boldsymbol T}\circ {\boldsymbol S}(ct,{\boldsymbol x})$,
where ${\boldsymbol S}$ is
a transformation $(w', {\boldsymbol x}')= {\boldsymbol\Psi}^{-1}(w,{\boldsymbol x})$
that preserves the independent variable and ${\boldsymbol T}$ is a time redefinition
$dt' = \Lambda(w,{\boldsymbol x}) dt$. (Another advantage is that the 
analysis extends immediately to ${\boldsymbol F}$ with explicit time-periodic 
dependence.)
The solutions of (\ref{DSC}) are unbounded along the $w$-direction, but this is not 
a problem since the nonzero LEs of system (\ref{DSC})
are identical to those of (\ref{DS}).

There is a caveat, however: the {\it spatial} boundedness of the solutions is not
preserved under Lorentz transformations.
A trajectory confined
to a bounded space-like region  ($\sup ||{\boldsymbol x}(t)||<\infty)$
of the first reference frame  
is seen as spatially unbounded from the other inertial reference frame. 
Similar problem is observed even for Galilean transformations, but in classical dynamics
one can adopt a reference frame where the solutions are bounded. In relativistic dynamics
such a choice would raise questions about the invariance of the LEs,
which is precisely the object of this Letter.

To proceed we first make the crucial observation that the study of chaos {\it can} be extended 
to this class of spatially
unbounded orbits, even though the same
does
not hold true for unbounded systems in general. Indeed,
sensitive dependence on initial conditions and LEs depend exclusively on the relative time
evolution between nearby trajectories; their dependence on the reference frame 
is limited to the definition
of the spacetime coordinates used to measure the
distances between the neighboring trajectories 
as they evolve over {\it identical} time intervals.
Therefore, chaos can
be properly defined and LEs can be used as indicators of chaos on an unbounded
trajectory
${\boldsymbol y}(t)$  insofar as {\it $||{\boldsymbol y}(t)-\hat{\boldsymbol y}(t)||$ remains
uniformly upper bounded for all $t$ and all trajectories $\hat{\boldsymbol y}(t)$ with initial
conditions in a neighborhood of ${\boldsymbol y}(0)$.}
That is, our condition is that the evolution of a small ball of points 
will remain bounded with respect to
the local observers at position ${\boldsymbol y}(t)$,
regardless of whether it remains bounded with respect to 
a fixed point of the reference frame. 
We refer to this as the {\it extended boundedness} condition. Note that this condition is satisfied
for ${\boldsymbol y}(t)$ interpreted as the extended coordinates ($w'(t), {\boldsymbol x}'(t)$)
after the transformation ${\boldsymbol S}$ whenever
the original system (\ref{DS}) is spatially bounded.

Having shown that LEs remain valid indicators of chaos despite the spatial unboundedness
of the transformed orbits, we now turn to the effect of the Lorentz  transformations on
the LEs. For the transformation ${\boldsymbol T}$, from  Eq. (\ref{Lorentz}) we have
\beq
\label{rep}
{dt'} = \gamma\left( 1 - \frac{v}{c^2}  F_x(\boldsymbol x(t)) \right){dt} \equiv
\Lambda({\boldsymbol x}(t))dt,
\eeq
where $ F_x(\boldsymbol x)$ stands for the $x$-component of $\boldsymbol  F(\boldsymbol x)$.
For $|F_x(\boldsymbol x(t))|\le c$,
implying $\inf \Lambda({\boldsymbol x}(t))>0$ in
the present case,
we have
$0 < \langle\Lambda \rangle_t = \lim_{t\rightarrow\infty} \frac{t'(t)}{t} =
\lim_{t\rightarrow\infty}\frac{1}{t} \int_0^t
\Lambda({\boldsymbol x}(p)) dp < \infty$ \cite{comment}.
This allows us to factor the LEs transformed by ${\boldsymbol T}\circ {\boldsymbol S}$ as
\begin{equation}
\tilde{\lambda}_i^{t'}  =  \tilde{\lambda}_i^{t}/\langle\Lambda \rangle_t,
\label{split}
\end{equation}
where $\langle\Lambda \rangle_t$ is the contribution due to ${\boldsymbol T}$
and $\tilde{\lambda}_i^{t}$ corresponds to $\lambda_i^{t}$ transformed by ${\boldsymbol S}$.
The problem is thus reduced to the transformation of the LEs under the
spatial transformation ${\boldsymbol S}$.
The nonsingular nature of (\ref{Lorentz}) assures the existence of the constants
$L^\pm$ necessary to establish  the bounds in (\ref{ge}) because, irrespective of the
spatial unboundedness,  the Jacobian matrix of the transformation 
is bounded.
Employing the Euclidean norm to the matrix
$D_{\boldsymbol y}{\boldsymbol\Psi}({\boldsymbol y})$ of (\ref{Lorentz}), we obtain
$ L^+ =L^- =\sqrt{(c+|v|)/(c-|v|)}$, leading to $\tilde{\lambda}_i^{t}  = {\lambda}_i^{t}$.
In particular, all positive LEs remain positive under this transformation. 

Combined with Eq.~(\ref{split}), this results in
$ {\tilde\lambda}_i^{t'} =  {\lambda}_i^{t}/{\langle\Lambda \rangle_t}$, which
is precisely the transformation (\ref{TLE}) previously established for the case 
of bounded orbits \cite{MotterPRL}.  Our result does not agree with the result
presented in~\cite{Zheng}
for 
averages over local LEs~\cite{local}, but that is because 
that study was restricted to time dilatations and length contractions, which correspond 
to the transformation of dynamical variables such as volume (or the reciprocal of a density) 
for the time 
measured at a fixed point of the reference frame, whereas our analysis 
describes single-particle dynamics for the time measured at the position of the particle.

If system (\ref{DS}) involves the evolution of velocities, as expected for a particle
in a 3D potential, the Lorentz transformation (\ref{Lorentz}) must be extended to include
the transformation of ${\boldsymbol u}\equiv d{\boldsymbol x}/dt$ into ${\boldsymbol u}'\equiv d{\boldsymbol x}'/dt'$,
which is given by
$u'_x=\eta(u_x-v)$,
$u'_y=\eta\gamma^{-1} u_y$,
and 
$u'_z=\eta\gamma^{-1} u_z$,
where $\eta=1/(1-u_xv/c^2)$.
The resulting transformation
$(w,{\boldsymbol x},{\boldsymbol u})\rightarrow (w',{\boldsymbol x}',{\boldsymbol u}')$
satisfies the extended boundedness condition and has constants $0<L^\pm<\infty$, as long as $|v|<c$
and $|u_x(t)| \le c$. This ensures that the LEs of systems obtained by order reduction of 
second-order differential equations, which are the most common in particle dynamics, will 
be transformed as in (\ref{TLE}) under Lorentz transformations.

\begin{figure}[t]
\resizebox{0.45\linewidth}{!}{\includegraphics*{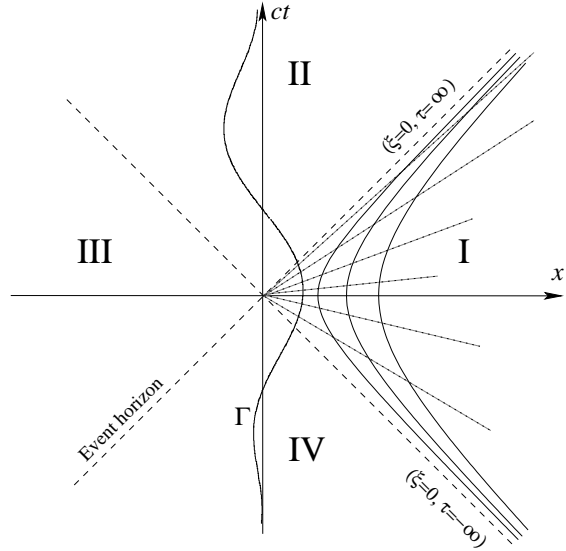}}
\caption{Accessibility to the dynamics is observer dependent.
The lines of fixed $\xi$ in Rindler coordinates correspond to
hyperbolic trajectories in the coordinates $(ct,x)$ of
inertial observers. 
The straight dotted
lines are the lines of constant time $\tau$. 
The uniformly accelerated observers are
unaware of all events occurring in regions II and III
of the original Minkowski spacetime.
They only access the dynamics of 
a trajectory $\Gamma$ during the time the trajectory crosses region I \cite{remote}. 
If the trajectory is spatially bounded with respect to the original observers,
as assumed for system (\ref{DS}), this corresponds to an infinite
time interval $\Delta\tau$ but only
to a finite time interval $\Delta t$. 
}
\label{fig2}
\end{figure}

{\bf Rindler transformations.}
With respect to an inertial reference frame, an observer with constant proper
acceleration $a$ along the $x$ direction has a hyperbolic worldline given by
\beq
\label{hyper}
ct(\tau) = \frac{c^2}{a} \sinh \frac{a\tau}{c}, \quad
x(\tau) = \frac{c^2}{a} \cosh \frac{a\tau}{c},
\eeq
where $\tau$ stands for the observer's proper time.
The corresponding Rindler transformation~\cite{horizon}
is defined by $(ct,x,y,z) \rightarrow (c\tau(t,x),\xi(t,x),y,z)$, with
\beq
\label{rindler}
ct(\tau,\xi) = c\sqrt{\frac{2\xi}{a}}\sinh \frac{a\tau}{c}, \;
x(\tau,\xi) = c\sqrt{\frac{2\xi}{a}}\cosh \frac{a\tau}{c},
\eeq
and positive $\xi$ (see Fig. \ref{fig2}).
The observer on hyperbole (\ref{hyper}) is in the Rindler reference frame
at rest at $\xi = c^2/2a$. In contrast with the Lorentz case, the Rindler
transformations are non-linear in $x$ and $ct$.

Focusing on the space defined by the extended configuration-space coordinates,
the matrix $D_{\boldsymbol y}{\boldsymbol\Psi}({\boldsymbol y})$ and its inverse
for the Rindler transformation of (\ref{DSC}) have unit determinants but their largest
eigenvalues diverge as $ c/\sqrt{2a\xi}\cosh a\tau/c$ for $\xi\rightarrow 0$.
Therefore, one cannot identify finite constants $L^\pm$ that could be used to compare
$\tilde\lambda^t_i$ and $ \lambda^t_i$. This behavior can be interpreted
in terms of our extended boundedness condition, which is not satisfied in this
case because $(c\tau(t,x(t)),\xi(t,x(t)),y(t),z(t))$ diverges at the light cone and 
is undefined beyond it. 
Moreover, from the inverse of Eqs. (\ref{rindler}), 
we have
\begin{eqnarray}
\label{rep1}
{d\tau} = \frac{c^2}{a}
\left[
\frac{x(t)- t F_x({\boldsymbol x}(t))}{x(t)^2- (ct)^2}\right] dt \equiv
\Lambda(ct,{\boldsymbol x}(t)) dt,
\end{eqnarray}
where $\Lambda(ct,{\boldsymbol x}(t))$ diverges when the original solution
$(ct, {\boldsymbol x}(t))$ crosses the light cone $x^2 = c^2t^2$. The same 
holds true for the physical time $dt'=\sqrt{2a\xi/c^2}d\tau$. 
The average $\langle \Lambda\rangle_t$ is not well defined and, as a result, 
the Rindler transformed system does not have a natural probability measure against 
which the LEs could be calculated~\cite{MotterPRL}. 
Therefore, the question of how the LEs transform under Rindler transformations
is ill-posed.

The real origin of the problem is the horizon structure (and its counterpart structure
for $t\rightarrow -t$) inherent to uniformly accelerated observers~\cite{horizon}. The 
Rindler transformation (\ref{rindler}) is not a global spacetime diffeomorphism since 
it maps only one quarter of the Minkowski spacetime, as shown in Fig.~\ref{fig2}.
Any event located above the component of the light cone corresponding to the bisectrix in the
first and third quadrants of Minkowski spacetime will never reach the accelerated observers. 
While singularities can be an artifact of the coordinates, event horizons are an attribute of the
reference frame. The existence of an
event horizon prevents the observers from having access to the
asymptotic dynamics of the original system. Therefore, without having access to the complete dynamics,
the Rindler observers cannot formulate a criterion for chaotic behavior---based on the observation
of individual trajectories---that is valid for the original system \cite{remote}.
It is interesting to notice that such a  problem, related to the global structure of
the spacetime, manifests itself as a violation of our conditions for the transformation
of LEs.

If one insists on computing the LEs from a uniformly accelerated referential frame~\cite{Zheng}, one
must note that the late-time dynamics of the extremely dilated time $\tau\rightarrow\infty$ does not correspond
to the real late-time dynamics of the original system since the interval $-\infty<\tau<\infty$ is the
mapping of a finite time interval $\Delta t$. Therefore even if one could compute $\tilde\lambda^{\tau}_i$
as seen from the accelerated frame, this would be,  in fact,  a problem different from the originally
proposed one. This situation is analogous to the limits imposed by the cosmological singularity to
the determination of chaos in FRW cosmologies~\cite{MotterPRD} and is also predicted for
Rindler transformations of any other 
dynamical system and for any choice of coordinates.

{\bf Rotating frames.}
The crucial role played by the event horizon in
the Rindler case can be better appreciated if one considers 
a physical situation involving 
a non-linear transformation that does not
introduce event horizons. This is precisely the
case of uniformly rotating reference frames~\cite{rotating}: $r'=r$, $\theta'=\theta+\Omega t$, $z'=z$, and
$cdt'= [ g(r)  + \Omega^2 r^2/g(r)]dt + [\Omega r^2/g(r)]d\theta$, where $g(r)=\sqrt{c^2-\Omega^2r^2}$, $\Omega$ 
is a constant, and $t'$ is the physical time in the rotating frame~\cite{PhysicalTime}.
This leads to
\begin{equation}
dt'= \left( \frac{g(r(t))}{c}  + \frac{\Omega r^2(t)[\Omega + F_{\theta}({\boldsymbol x}(t))]}{cg(r(t))}   \right) dt,
\label{rot2}
\end{equation}
where $F_{\theta}({\boldsymbol x})=d\theta/dt$.
The transformation of the LEs  of (\ref{DSC}) is in this case well defined 
since the extended boundedness condition is satisfied 
for orbits in closed sets of the physical 
region $|\Omega| r < c$ for which  $-\Omega r^2 F_{\theta}({\boldsymbol x}) <c^2$,
where both the function $\Lambda({\boldsymbol x})$
and the constants $L^\pm$ are upper and lower bounded away
from zero. The latter follows from the fact that the 
entries of the Jacobian matrix
$D_{\boldsymbol y}{\boldsymbol\Psi}({\boldsymbol y})$  and its inverse for
the transformation $(ct,r,\theta,z)\rightarrow(ct',r',\theta',z')$ are all continuous for $\Omega r <c$.
A subtlety in this calculation is that in rotating frames the differential $dt'$ of the physical 
time is not exact and cannot be integrated globally, meaning that the Jacobian elements involving 
derivatives of $ct'$ must be determined from $cdt'$ in the immediate neighborhood of a given $r$. 
The transformation $t\rightarrow t'$ is defined locally but it can always be extended along any 
trajectory with initial condition in that neighborhood.
Therefore, the LEs transform as predicted by (\ref{TLE}) also for the case of rotating frames.

{\bf Generalization and discussion.}
Our derivation of Eq. (\ref{split}) also demonstrates that conditions ({\it i}-{\it iii}) are
sufficient (and usually necessary) for the validity of (\ref{TLE}) in general. 
Indeed, while we considered specific 
transformations and specific classes of dynamical systems in our explicit examples, 
these three conditions are 
precisely the {\it checkpoints} we have to verify for any system and any transformation. The
extended boundedness condition---satisfied both before {\it and} after the 
transformation in the {\it extended space}, which includes $ct$ as an additional
coordinate---guarantees that the system can be kept autonomous and that LEs remain
valid indicators of chaos. The condition that
the Jacobian is bounded---in the sense of having positive finite constants $L^{\pm}$ for 
the transformation in the {\it extended space}---ensures 
the validity of the identity $\tilde{\lambda}_i^{t}  = {\lambda}_i^{t}$. Finally, $\Lambda$ 
and $\langle\Lambda\rangle_t$ positive and finite---again, in the {\it extended space}---guarantees 
that the time transformation is well defined and the signs of the LEs are conserved; it also 
guarantees that the time transformation is invertible, a condition we saw violated for the Rindler transformation.

These conditions are readily applicable to {\it any} system and {\it any} change of reference
frame and coordinates. The latter includes the choice of the time parameter or of the 
observers in the reference frame with respect to which the time is measured.
In the examples above, the dynamical system describes the dynamics of a single particle, the 
dynamical variables represents the coordinates and possibly velocities of the particle, and
the time was assumed to be recorded locally---each time by the observer in the reference frame 
that is at the point where the particle is. 
However, other choices are equally valid. For a many-particle system 
under Lorentz transformation, for example, the time could be measured, e.g., 
with respect to the position of one of the particles, 
${dt'} = \gamma\left( 1 - \frac{v}{c^2}  F_{x_i}(\boldsymbol x(t)) \right){dt}$,
with respect to the center of mass, 
${dt'} = \gamma\left( 1 - \frac{v}{c^2} \sum_i \frac{m_i}{\sum_j m_j} F_{x_i}(\boldsymbol x(t) ) \right){dt}$, 
or with respect to a fixed point, 
${dt'} = \gamma {dt}$.
Moreover, the dynamical system can describe physical, chemical or biological 
activity whose dynamical variables do not necessarily correspond to coordinates and velocities in 
the physical space. In this general case the system can be written as 
$\frac{d}{dt}\mbox{X}_i=F_i(\mbox{X}_1, \dots \mbox{X}_n)$,  
$i=1,\dots n$, and the transformation is locally defined  as 
$(cdt,d\mbox{X}_1,\dots d\mbox{X}_n)\rightarrow (cdt',d\mbox{X}'_1,\dots d\mbox{X}'_n)$.
The latter is determined by the change of reference frame and spacetime coordinates, 
$(cdt, d\boldsymbol x) \rightarrow (cdt', d\boldsymbol x')$, and depends on the nature of the dynamical 
variables, i.e., whether they transform as scalars, vectors, tensors, or in a different way. 
The choice of observers in the new reference frame is always accounted for 
through the choice of $\frac{d}{dt}{\boldsymbol x}$
in the transformation formula 
$dt'=\left(\frac{\partial}{\partial t} t'(ct,\boldsymbol x) + \nabla_{\boldsymbol x}  t'(ct,\boldsymbol x)\cdot \frac{d}{dt}\boldsymbol x \right)dt$, where this term vanishes only if the time is measured (remotely) by a fixed observer.

The results presented in this Letter address all these cases and show that,
if conditions ({\it i-iii}) are verified, the signs of the LEs remain valid 
invariant indicators of chaos. 
Since we have extended the use of the LEs as a valid measure of chaos to include unbounded orbits, 
this conclusion is general: it applies to both inertial and non-inertial reference frames and does not
involve the identification of privileged observers.  
These results account for properties inherent to relativistic observers, such as event
horizon and spatial unboundedness, significantly extending our understanding of the
relativistic invariance of LEs and chaos.

The authors thank E. Gu\'eron, G. Matsas and R. Venegeroles for insightful discussions.
This work was supported by FAPESP and CNPq.


\begin{thebibliography}{99}

\bibitem{ChaosGR}
G. Francisco and G.\ E.\ A. Matsas, Gen. Relativ. Gravit. {\bf 20}, 1047 (1988). For an early
review, see
D. Hobill, A. Burd, and A. Coley (Eds.), {\em Deterministic Chaos in General Relativity} (Plenum
Press, New York, 1994).

\bibitem{Cipr_Hor}
P. Cipriani and M. Di Bari, Phys. Rev. Lett. {\bf 81}, 5532 (1998);
L. Horwitz {\it et al.}, {\it ibid} {\bf 98}, 234301 (2007).

\bibitem{MotterPRL} A.\ E. Motter, Phys. Rev. Lett. {\bf 91}, 231101 (2003).

\bibitem{Ott} E. Ott, {\em Chaos in Dynamical Systems} (Cambridge Univ. Press, Cambridge, 1994).

\bibitem{Oseledec} V.\ I. Oseledec, Trans. Moscow Math. Soc. {\bf 19}, 197 (1968).


\bibitem{Jaroslawski} 
R. Jaroslawski {\it et al.},
Z. Phys. B {\bf 82}, 437 (1991).

\bibitem{Eichhorn} R. Eichhorn, S.\ J. Linz, and P.\ H\"anggi, Chaos, Solitons \& Fractals {\bf 12}, 1377 (2001).


\bibitem{MotterPRD} A.\ E. Motter and P.\ S. Letelier, Phys. Rev. D {\bf 65}, 068502 (2002).

\bibitem{Zheng} Z. Zheng, B. Misra, and H. Atmanspacher, Int. J. Theor. Phys. {\bf 42}, 869 (2003).


\bibitem{comment}
Similarly, $0  < \langle\Lambda \rangle_t^{-1} =
\langle\Lambda^{-1} \rangle_{t'} = \lim_{t'\rightarrow\infty} \frac{t(t')}{t'}  < \infty$.

\bibitem{local}
B. Eckhardt and D. Yao, Physica D {\bf 65}, 100 (1993);
S. V. Ershov and A. B. Potapov, {\it ibid} {\bf 118}, 167 (1998);
P. Gaspard, {\it Chaos, Scattering and Statistical Physics} (Cambridge Univ. Press, Cambridge, 1998).



\bibitem{horizon} C. de Almeida and  A. Saa,   Am. J. Phys. {\bf 74}, 154 (2006).

\bibitem{remote}
In principle, they can remotely observe the dynamics for $t<0$
and determine the backward LEs based on that information.

\bibitem{rotating} J.\ R. Letaw and J.\ D. Pfautsch, Phys. Rev. D {\bf 22}, 1345 (1980);
                   J. Math. Phys. {\bf 23}, 425 (1982).
                   
\bibitem{PhysicalTime} R.\ J. Cook, Am. J. Phys. {\bf 72}, 214 (2004).



\end{thebibliography}
\end{document}